Research

# Comparative analysis of transposed element insertion within human and mouse genomes reveals *Alu*'s unique role in shaping the human transcriptome

Noa Sela*, Britta Mersch†, Nurit Gal-Mark*, Galit Lev-Maor*, Agnes Hotz-Wagenblatt† and Gil Ast*

Addresses: *Department of Human Molecular Genetics and Biochemistry, Sackler Faculty of Medicine, Tel Aviv University, Ramat Aviv 69978, Israel. †HUSAR Bioinformatics Lab, Department of Molecular Biophysics, German Cancer Research Center (DKFZ), Im Neuenheimer Feld, D-69120 Heidelberg, Germany.

Correspondence: Gil Ast. Email: gilast@post.tau.ac.il

## Abstract

**Background:** Transposed elements (TEs) have a substantial impact on mammalian evolution and are involved in numerous genetic diseases. We compared the impact of TEs on the human transcriptome and the mouse transcriptome.

**Results:** We compiled a dataset of all TEs in the human and mouse genomes, identifying 3,932,058 and 3,122,416 TEs, respectively. We than extracted TEs located within human and mouse genes and, surprisingly, we found that 60% of TEs in both human and mouse are located in intronic sequences, even though introns comprise only 24% of the human genome. All TE families in both human and mouse can exonize. TE families that are shared between human and mouse exhibit the same percentage of TE exonization in the two species, but the exonization level of *Alu*, a primate-specific retroelement, is significantly greater than that of other TEs within the human genome, leading to a higher level of TE exonization in human than in mouse (1,824 exons compared with 506 exons, respectively). We detected a primate-specific mechanism for intron gain, in which *Alu* insertion into an exon creates a new intron located in the 3' untranslated region (termed 'intronization'). Finally, the insertion of TEs into the first and last exons of a gene is more frequent in human than in mouse, leading to longer exons in human.

**Conclusion:** Our findings reveal many effects of TEs on these two transcriptomes. These effects are substantially greater in human than in mouse, which is due to the presence of *Alu* elements in human.

## Background

The completion of the human and mouse genome draft sequences confirmed that transposed elements (TEs) play a major role in shaping mammalian genomes [1,2]. Transposed elements comprise at least 45% of the human and 37% of the mouse genomes. In the human genome, *Alu* is the most abundant transposed element (TE), comprising more than one million copies, which is about 10% of the genome. We previously reported that more than 5% of the alternatively spliced internal exons in the human genome are derived from *Alu*, and to the best of our knowledge all *Alu*-driven exons originated from exonization of intronic sequences [3,4]. *Alu* elements were shown to create alternative cassette exons, whereas exonization of a constitutively spliced exon was shown to have deleterious effects [4,5]. Alternatively spliced *Alu* exons thus enrich the transcriptome, the coding capacity, and the regulatory versatility of primate genomes with new isoforms, without compromising the integrity and the original repertoire of the transcriptome and its resulting proteome. Therefore, exonization with low inclusion level is thought to be the playground for future possible exaptation (adopting a new function that is different from its original one) [6] and fixation within the human transcriptome [3,7-11].

Several indications imply that *Alu* insertions can add new functionality to proteins, such as exon 8 of ADAR2 gene [12]. An analysis of protein databases indicates that mammalian interspersed repeat (MIR) and CR1 (chicken repeat 1) TEs can contribute to human protein diversification also [7]. Moreover, ultraconserved exons were found to originate from an old short interspersed nuclear element (SINE) [13]. Another important role for new exonizations is a potential tissue specificity, in which many minor form exons (which are mainly new exonizations) exhibit strong tissue regulation [14]. Experimental support for this bioinformatics analysis is given by a report of *Alu de novo* insertion and subsequent exonization within the dystrophin, creating a tissue-specific exon that results in cardiomyopathy [15]; *Alu* exonization within the *NARF* gene was also shown to differ among human tissues [16].

TEs are also thought to contribute to the turnover of intron sequences, because there is often equilibrium between sequence gain (by TEs) and sequence loss by unequal crossing over between TEs [17]. Sironi and coworkers [18] identified constraints on insertion of transposed elements within introns, and they showed that gene function and expression influence insertion and fixation of distinct transposon families in mammalian introns [19].

The origin of spliceosomal introns is a longstanding unresolved mystery. It was recently demonstrated that the duplication of small genomic portions containing 'AGGT' provides the boundaries for new introns [20]. In only two cases is the origin of the intron known: a SINE insertion that gave rise to a new intron in the coding region of the catalase A gene of rice, and two midge globin genes that acquired an intron via gene conversion with an intron-containing paralog [21,22]. It has been postulated that humans underwent only intron loss and not intron gain [23,24], and new introns that originated from SINE insertion have not been reported in vertebrates.

In addition to *Alu*, the human genome contains multiple copies of other families of TEs, including MIR (a tRNA-derived SINE) and long interspersed nuclear element (LINEs) such as (LINE)-1 (L1), LINE-2 (L2), and CR1 (L3). The mouse genome contains MIR elements as well as rodent-specific SINEs, such as B1, which is a 7SL RNA-derived TE that originated from the same ancestral sequence as the left arm of the *Alu*; B2, B4, and ID, which are tRNA-derived SINEs; and LINEs such as L1, L2, and CR1. The human and mouse genome also contain several copies of long terminal repeats (LTRs) and DNA repetitive elements. The latter were recently shown to be intensively active in the primate lineage [25]. The mouse genome was chosen for comparative analysis of TE insertions, because this genome contains a TE originating from the same ancestral sequence of the *Alu* (B1) [26] in multiple copies, as well as the fact that complete annotations of the genome are available, and there is a high coverage of the mouse transcriptome by expressed sequence tags (ESTs) and cDNAs.

In this work, we addressed several questions concerning the global effect of TEs on the human transcriptome and whether the exonization process is unique to primates or is shared by other mammals as well. More specifically, we wished to answer the following questions. Do all TE families exonize? Do all TEs have the same exonization rate? Are some of these newly created exons tissue-specific? Furthermore, inasmuch as cancerous tissues have been shown to adopt aberrant splicing patterns [27], are there TE exonizations that are potentially cancer specific? Can we detect exonized TEs that are not alternatively spliced? Are TE insertions responsible for the origin of new introns within the human or mouse genome? TEs are inserted into introns in sense and antisense orientations relative to the mRNA precursor. Hence, do exonized TEs have a preferential orientation, and how many of them contribute a whole exon? Do TEs enter into all parts of the mRNA with the same probability? How many of these exonizations potentially contribute to proteome diversity? And finally, do they possess the same characteristics as conserved alternatively spliced cassette exons?

To address these questions, we compiled a dataset of all SINE, LINE, LTR, and DNA TEs in the human and mouse genome. We analyzed insertions into introns and the effect of TE insertions on the transcriptome. Our analysis indicates that TEs have a greater effect on shaping the human transcriptome than the mouse transcriptome. This effect is 3.6 times greater in human than in mouse, and this is caused by a higher level of exonization of the *Alu* element, which is a primate-specific TE. Four lines of evidence support our finding. First, the exonization level of *Alu* is significantly greater compared with other TEs within the human transcriptome. Second, all TEs within the mouse transcriptome have the same exonization level. Third, TEs that belong to the same families, such as MIR, LINE-2, and CR1, exonize in the same level in both species. Finally, the level of TE exonization in human compared

**Table 1**

**TE effect on the human transcriptome**

| RE | Total | Intronic | TE in introns of UCSC annotated genes[a] | TE in introns of non-annotated genes[a] | TE exonization in UCSC annotated genes[a] | TE exonization in non-annotated genes[a] |
|---|---|---|---|---|---|---|
| *Alu* | 1,094,409 | 718,460 (66%) | 480,052 | 238,408 | 1060 (0.2%) | 584 (0.2%) |
| MIR | 537,730 | 351,366 (65%) | 231,893 | 119,473 | 181 (0.08%) | 134 (0.1%) |
| L1 | 830,062 | 486,901 (58%) | 282,146 | 204,755 | 219 (0.08%) | 250 (0.1%) |
| L2 | 375,116 | 240,350 (64%) | 154,309 | 86,041 | 103 (0.07%) | 72 (0.08%) |
| CR1 | 50,156 | 33,365 (66%) | 22,087 | 11,278 | 12 (0.04%) | 6 (0.05%) |
| LTR | 654,897 | 292,456 (44%) | 136,461 | 155,995 | 155 (0.1%) | 150 (0.09%) |
| DNA | 389,688 | 226489 (58%) | 145,968 | 80,521 | 93 (0.06%) | 142 (0.17%) |
| Total | 3,932,058 | 2,349,387 (60%) | 1,452,916 | 896,471 | 1824 (0.12%) | 1653 (0.18%) |

Insertions of transposed elements (TEs) within the human genome. The different classes of the examined TEs are shown in the left column. 'Total' (second column) indicates the overall amount of each TE within the human and mouse genomes. 'Intronic' (third column) indicates the number of TEs within intronic regions, and the percentage of TEs within introns relative to the total amount of TEs is shown in parentheses brackets. The fourth and fifth columns show the number of TEs within introns of the University of California, Santa Cruz (UCSC) knownGene list (version hg17) and those inserted within genes not listed within UCSC knownGene list. The sixth and seventh columns show the number of exonized TEs within the UCSC knownGene list and those inserted within genes not listed within UCSC knownGene list. In parentheses are indicated the percentage of exonized TEs is indicated. The lower row shows the total number of all TEs. [a]Gene annotation is based on the annotations of the known gene list in the UCSC genome browser (version hg17). LTR, long terminal repeat; MIR, mammalian interspersed repeat; RE, retroelement.

with mouse is significantly greater after normalization for differences in transcript coverage. Moreover, we found that *Alu* insertion within exons in the human transcriptome, a process termed 'intronization', creates a new alternative intron, which is a primate-specific intron of the intron retention type. Finally, these findings indicate that *Alu* elements play many important roles in shaping human evolution, presumably leading to a greater degree of transcriptomic complexity.

## Results
### Genome-wide survey of transcripts containing transposed elements

To evaluate the effect of TEs on the human and mouse transcriptome, we calculated the total number of TEs in both genomes, the number of TEs in introns, and the number of TEs that are present within mRNA molecules. We therefore downloaded EST and cDNA alignments, as well as repetitive elements' annotations of the human genome and the mouse genome from the University of California, Santa Cruz (UCSC) genome browser (hg17 and mm6, respectively) [28], and analyzed for TE insertions (see Materials and methods, below). Our analysis of the numbers of TEs in the human and mouse genomes is summarized in Tables 1 and 2, respectively. There are approximately 3.9 and 3.1 million copies of TEs in the human and mouse genomes, respectively. The most abundant TE families within the human genome are *Alu* and L1 elements, with almost 1.1 million and 800,000 copies each. The most abundant TE families in the mouse genome are L1 (800,000 copies) and B1 (500,000 copies).

Next, we examined the number of TEs in introns. It is interesting to note that all families of TEs have a tendency to reside within intronic regions. Between 44% and 66% of TE inser-tions are located within intronic sequences. *Alu* in humans and B4 in mice have the highest ratio of insertions within introns (66%), whereas L1 and LTR both in human and mouse have the lowest percentage of copies within introns (58% in human and 56% in mouse for L1, 44% in human and 52% in mouse for LTR). L1 and LTR exhibit a biased insertion in the antisense orientation relative to the mRNA within intronic sequences in both human and mouse: 185,428 and 96,718 L1 repeats were inserted in the antisense and sense orientations in human, respectively; 113,862 and 68,101 L1 repeats in mouse; 96,654 and 39,804 LTRs in human; and 101,001 and 55,689 LTRs in mouse. No such bias was detected in SINEs, or in L2, CR1, and DNA repeats. This shows a tendency toward insertion or fixation of all TEs into intronic sequences.

### Did all transposed elements families undergo exonization, and do they all have the same exonization level?

TEs present in EST/cDNA were separated into those that were entered within annotated genes (according to the knownGene list in UCSC; see Materials and methods, below) and those that were not mapped to known genes. These were considered non-protein-coding genes (see Materials and methods, below).

We then examined exonization of TEs, that is, an internal exon in which a TE is either as part of or as the entire exon sequence. All TE families in both human and mouse can undergo exonization (Tables 1 and 2, respectively; the two right-most columns). We found a much higher level of TE exonization in the human transcriptome than in the mouse transcriptome. We calculated the exonization level (LE) as the percentage of TEs that exonized within the number of

**Table 2**

**TE effect on the mouse transcriptome**

| RE | Total | Intronic | TE in introns of UCSC annotated genes[a] | TE in introns of non-annotated genes[a] | TE exonization in UCSC annotated genes[a] | TE exonization in non-annotated genes[a] |
|---|---|---|---|---|---|---|
| B1 | 506,528 | 331,015 (65%) | 189,268 | 141,747 | 134 (0.07%) | 96 (0.07%) |
| MIR | 116,355 | 66,597 (63%) | 41,853 | 24,744 | 27 (0.06%) | 14 (0.06%) |
| B2 | 338,642 | 215,264 (63%) | 118,646 | 96,618 | 81 (0.07%) | 80 (0.08%) |
| B4 | 345,646 | 216,550 (66%) | 119,827 | 96,723 | 62 (0.05%) | 72 (0.07%) |
| ID | 45,955 | 30,285 (57%) | 18,022 | 12,263 | 8 (0.04%) | 3 (0.02%) |
| L1 | 820,434 | 457,705 (56%) | 181,292 | 276,413 | 102 (0.07%) | 189 (0.07%) |
| L2 | 56,518 | 34,923 (62%) | 18,963 | 15,960 | 9 (0.05%) | 5 (0.03%) |
| CR1 | 11,812 | 7,167 (61%) | 3,779 | 3,388 | 0 (0%) | 1 (0.03%) |
| LTR | 756,324 | 396,226 (52%) | 156,690 | 239,536 | 72 (0.05%) | 243 (0.1%) |
| DNA | 124,202 | 75,200 (60%) | 40,428 | 34,772 | 11 (0.02%) | 19 (0.05%) |
| Total | 3,122,416 | 1,830,932 (58%) | 888,768 | 942,164 | 506 (0.06%) | 722 (0.08%) |

Insertions of transposed elements (TEs) within the mouse genome. The different classes of the examined TEs are shown in the left column. 'Total' (second column) indicates the overall amount of each TE within the human and mouse genomes. 'Intronic' (third column) indicates the number of TEs within intronic regions, and the percentage of TEs within introns relative to the total amount of TEs is shown in parentheses. The fourth and fifth columns show the number of TEs within introns of University of California, Santa Cruz (UCSC) knownGene list (version mm6) and those inserted within genes not listed within UCSC knownGene list. The sixth and seventh columns show the numbers of exonized TEs within the UCSC knownGene list and those exonized within genes not listed within UCSC knownGene list. In brackets are indicated the percentage of exonized TEs. The lower row shows the total number of all TEs. [a]Gene annotation is based on the annotations of the known gene list in the UCSC genome browser (version hg17). LTR, long terminal repeat; MIR, mammalian interspersed repeat; RE, retroelement.

intronic TEs (also see Materials and methods, below). In humans, 0.12% of the TEs exonized within protein coding genes (1,824 TE exonizations out of 1,452,916 TEs in introns) and 0.18% of the TEs exonized within non-protein-coding genes (1,653 out of 896,471). In contrast, we found a 0.06% rate of exonization within protein coding genes (506 out of 888,768) and 0.08% (722 out of 942,164) in non-protein-coding genes in the mouse transcriptome. The higher level of exonization in human compared with that in mouse is significant even after normalization of the relative EST/cDNA coverage (7.9 million transcripts in human versus 4.7 million transcripts in mouse - a ratio of 1.7). That is, even if we multiply the exonization of mouse by 1.7, there is still significantly higher exonization in the human genome ($\chi^2$ Fisher's exact test; $P < 10^{-29}$ [degrees of freedom = 1] for protein-coding genes and $P < 10^{-19}$ [degrees of freedom = 1] for non-protein-coding genes, for a multiplication by 1.7 of the exonization level within the mouse genome).

When the dataset was further reduced to exons in which there were at least two ESTs/cDNAs, confirming their exonization, we also observed a higher exonization level within human genome: 0.05% exonization in human both in coding and non-protein-coding genes, versus 0.03% and 0.02% in mouse coding and non-protein-coding genes, respectively ($\chi^2$; $P < 10^{-16}$ [degrees of freedom = 1] for protein-coding genes and $P < 10^{-22}$ [degrees of freedom = 1] for non-protein-coding genes; see Additional data file 1). The importance of long non-protein-coding RNA was recently demonstrated in human transcripts [29]. We therefore present an example of an exonization within a non-protein-coding gene (Additional data file 5). The fact that more than 50% of our data are supported by only one item of EST/cDNA evidence raises questions regarding the fidelity of the spliceosome (see Discussion, below).

Several TE families are located in the human and mouse genome, including MIR, L1, L2, CR1 (L3), LTR, and DNA repeats; thus, we can expect there to be a substantial amount of orthologous TE exons (exonization of the same TE in the human-mouse ortholog gene) in these families. However, only six TE exons were found to be orthologous, of which four are exonizations of MIR elements and two are exonizations of DNA repeats. It is doubtful that these are two independent insertion events because MIR and DNA repeats were active in common ancestors of all mammals, and because independent insertion into precisely the same locus is very rare. We therefore suggest that these MIR and DNA repeats were inserted into a common mammalian ancestor. These exons could either result from independent exaptation in the separated lineages or occur as a result of one exaptation event in the human-mouse common ancestor.

Do all TEs have the same exonization potential? That is, do all intronic TEs exhibit the same probability for acquiring mutations that subsequently lead the splicing machinery to select them as internal exons? Our analysis reveals that the majority of TE families exhibit similar exonization capabilities, at around 0.07% in both human and mouse (meaning 0.07% of the intronic TEs exonized). Statistical analysis indicated that there was no difference in the level of exonization of MIR, L1, L2, and CR1 and DNA within the human genome ($\chi^2 = 5.25$; $P$

= 0.26 [degrees of freedom = 4]), although LTR exonization in human was higher, compared with that of other SINEs, LINEs, and DNA repeats, but still substantially lower than *Alu*. Also, there were also no differences in exonization level between B1, B2, B4, ID, MIR, L1, L2, and CR1 within the mouse genome ($\chi^2$ = 10; $P$ = 0.18 [degrees of freedom = 7]), and LTR and DNA exhibited a slightly lower level of exonization in mouse. An exceptional case was the *Alu* exonization level, which was almost three times higher than that of all other TE families, with more than 0.2% of its intronic copies being exonized (all $\chi^2$ test values are listed in Additional data file 2). In addition, no differences were found in exonization level between the human and mouse MIR element, L2, and CR1. Interestingly, L1 exonization levels were higher in human than in mouse, and there was also a higher exonization level of LTR and DNA repeats in human compared with mouse. However, the L1 populations were different between human and mouse genomes (Additional data file 7), and the LTR and DNA populations were very heterogeneous. The LTR of the mouse was very abundant with the younger retroviral class II (ERVK), in which almost no exonization was detected.

In summary, these findings indicate that the *Alu* sequence is a better substrate for the exonization process, as compared with all other TE families. The higher level of exonization for *Alu* could be due to many 'unproductive' *Alu* exonizations, which were 'weeded out' in older exonizations. However, our comparison of TE families that were inserted into the genome at around the same time as *Alu* (L1 in human and B1, B2, and B4 in mouse) and which exhibited a much lower level of exonization than that of *Alu* probably indicates that *Alu* is a much better sequence for the exonization process than the others.

## Do transposed element exonizations have tissue specificity and cancer characteristics?
To examine TE exons that may be spliced differently among tissues, we used a bioinformatics analysis approach developed previously to identify tissue-specific exons [30]. We found 74 exons in human and 18 exons in mouse that putatively undergo tissue-specific splicing. In human, 41 exons belong to *Alu*, seven are MIR exons, seven are L1 exons, two are L2 exons, one is a CR1 exon, ten are LTR exons, and seven are DNA exons. In mouse, five are B1 exons, four are MIR exons, one is a B4 exon, one is an L1 exon, one is an L2 exon, and six LTR exons. (All of these exons are listed in Additional data file 13; the SINE, LINE, LTR, and DNA exons with tissue specificity score above 95 are listed in Additional data file 10 (parts B and C).

A bioinformatics approach to identifying exons that changed their splicing regulation in cancer is described by Xu and Lee [31]. We used this approach to analyze our data. We identified 36 such exons in human and 10 in mouse (listed in Additional data file 13). We further filtered our data to search for exons that were intronic within normal tissues and recognized as exons only within cancerous tissues and hence can serve as a potential marker for cancer diagnostics. Six such exons were found in six different genes (*ACAD9*, *YY1AP*, *KUB3*, *AMPK*, *NEL-like 1* and *active BCR-related* gene) and all of them were primate-specific *Alu* exons (Additional data file 10 [part A]). All exons were found within the coding sequence (CDS): in the *YY1AP*, *NEL-like1* and *active BCR-related* gene they introduce a stop codon, whereas in *ACAD9* and *KUB3* they cause frame shifts. It was only the *Alu* exon in *AMPK* that did not have a deleterious effect on the protein (it did not introduce a stop codon or cause a frame shift) and was not found to introduce a known protein domain. Except for the exonization within the *NEL-like-1* gene in which the isoform skipping the *Alu* exon (meaning the ancestral isoform) could not be detected within cancerous tissues, in all other genes the ancestral isoform was present within the cancerous tissue as well, probably only leading to reduction in the ancestral isoform concentrations. In one of these genes, namely *ACAD9*, we experimentally observed exonization in two ovarian cancer cell lines, but not in mRNA extracted from seven nonovarian cell lines (Additional data file 12).

## Can we detect exonized transposed elements that are not alternatively spliced?
The 1,824 human and 506 mouse TE exons can affect the transcriptomes in many different ways. In our data, 94% of the exonizations in human and 88% of the exonizations in mouse generated an internal cassette exon (Figure 1a [ii]; as was also reported elsewhere [3-5]). In the rest of the cases, the exonization formed alternative 5' splice sites (5'ss), alternative 3' splice sites (3'ss), or constitutively spliced exons. The numbers of the different splice forms of the TE exons in human and mouse are shown in Figure 1a. In the majority of cases, the alternative 5'ss or 3'ss is generated when an exon is alternatively elongated as a result of an alternative 5'ss or 3'ss selection within the TE (Figure 1a [iii] and 1a [iv], respectively). Also, in 3.1% and 5.7% of the human and mouse TE exonizations, respectively, the exons are detected *in silico* as constitutively spliced. In most of these cases (71%) the constitutively spliced exons were found in the untranslated region (UTR), and in 12.2% of the cases the constitutively spliced exon entered within the CDS and is 'divisible by 3' (preserve the reading frame, also termed symmetrical). In the rest of the cases, when the exonization is within the CDS and is not 'divisible by 3', the gene encodes a hypothetical protein.

Exon 2 of the *DMWD* gene originated from exonization of a MIR element. This exon is highly conserved within the mammalian class. Figure 2a,b show the alignments of the exon among human, chimpanzee, rhesus, mouse, rat, dog, and cow ortholog. The divergence of that exon, relative to the consensus MIR sequence, is high (about 25%). However, following exonization the exon is highly conserved among the species. This implies that once the exon has undergone exaptation and acquired a function, a purifying selection prevents accumulation of mutations. The high level of protein conservation

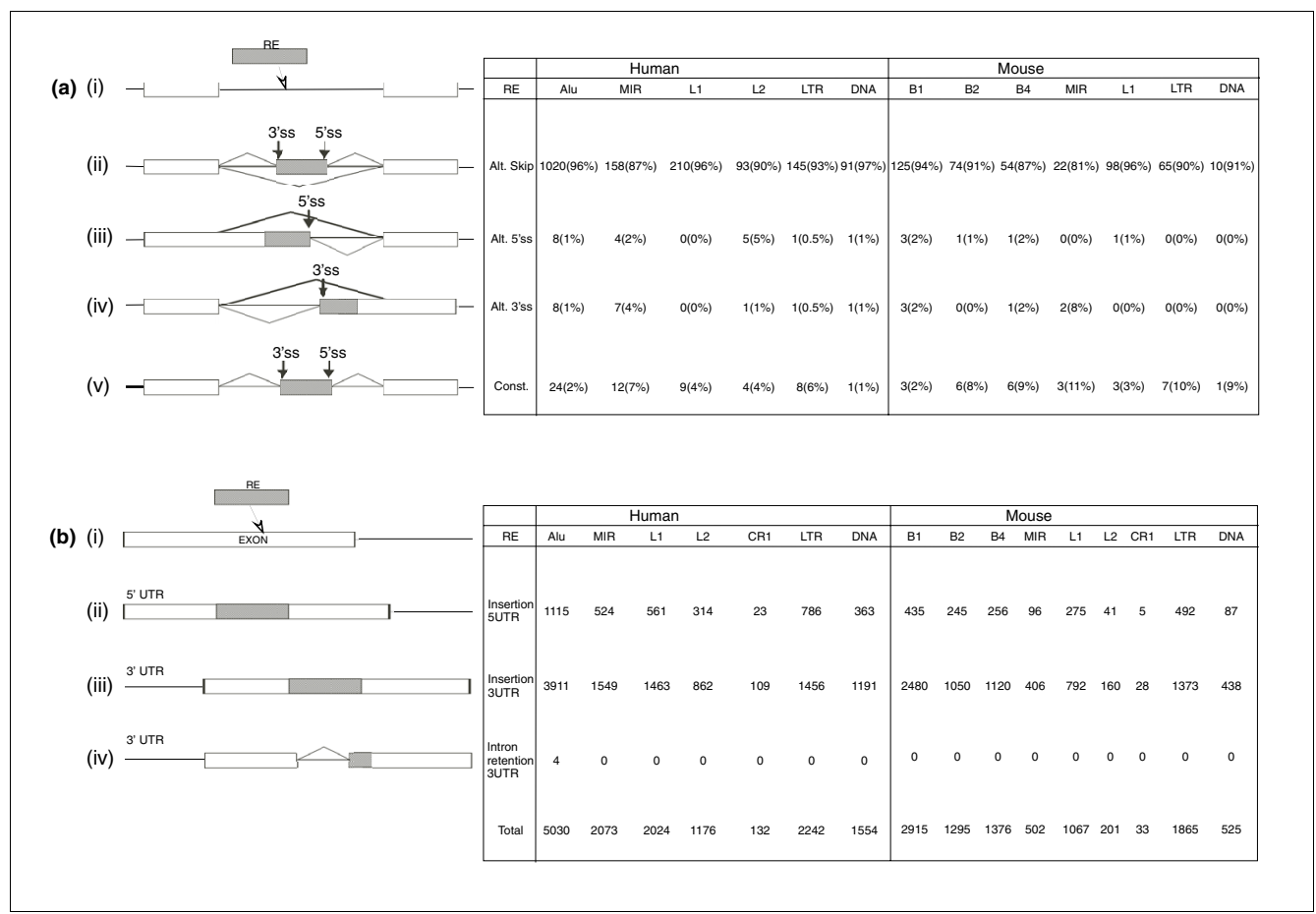

**Figure 1**
How TEs affect the human and mouse transcriptome. **(a)** Summary of the effect of (i) exonization of TEs on the transcriptome; of the effect of exonization that (ii) creates an alternatively skipped exon, (iii) transforms an existing exon to an alternative 5'ss exon, or (vi) transforms an existing exon to an alternative 3'ss exon; or of the effect of exonization that (v) creates a constitutively spliced exon. The table on the right shows the corresponding numbers of transposed elements (TEs). **(b)** Summary of the effect of TE insertions in the first or last exon. Panel i shows the insertion of TEs (gray box) into an exon (white box). The insertion of the TEs can cause an enlargement of the first or last exon (panels ii and iii) or, in some cases, activates intronization (generating an alternatively spliced intron that splits the last exon into two smaller exons; panel iv). The numbers of those events according to TE family are shown on the right-hand side.

(Figure 2b) suggests that exaptation occurred before the human, mouse, rat, dog, and cow split.

From the four MIR orthologous exons, two were selected for experimental validation. One was selected to show the conserved alternative splicing pattern between human and mouse, and the other to show the conserved constitutively spliced pattern between human and mouse. The *Alu* was chosen randomly from all constitutively spliced *Alu* exons found in our analysis. Figure 2c shows the validation of the splicing pattern of three exons. The first exon originating from MIR is conserved between human and mouse, and is alternatively spliced in both species (exon 2 of *DMWD* gene; Figure 2c, lanes 1 and 2); the second also originates from MIR, and is conserved between human and mouse, but it is constitutively spliced (exon 5 of *MYT1L* gene; Figure 2c, lanes 3 and 4); and the third one is an *Alu* exon, which is constitutively spliced

(exon 3 of *FAM55C* gene; Figure 3c, lane 5). This reverse transcription polymerase chain reaction (RT-PCR) analysis confirms that, under the above conditions and within the examined tissues, we can detect only one isoform that contains the exonization. This observation cannot exclude the possibility that this exon is alternatively spliced within other tissues or under different conditions.

## Transposed element insertion into last and first exons of the untranslated region

Furthermore, our analysis shows that the influence of TEs on the transcriptome is not limited to the creation of new internal exons from intronic TEs (exonization); TEs can also modify the mRNA, by being inserted within the first or last exon of a gene. The insertion causes an elongation of the first/last exons that are usually part of the UTR or an activation of an alternative intron (termed intronization; Figure 1b [ii to iv],

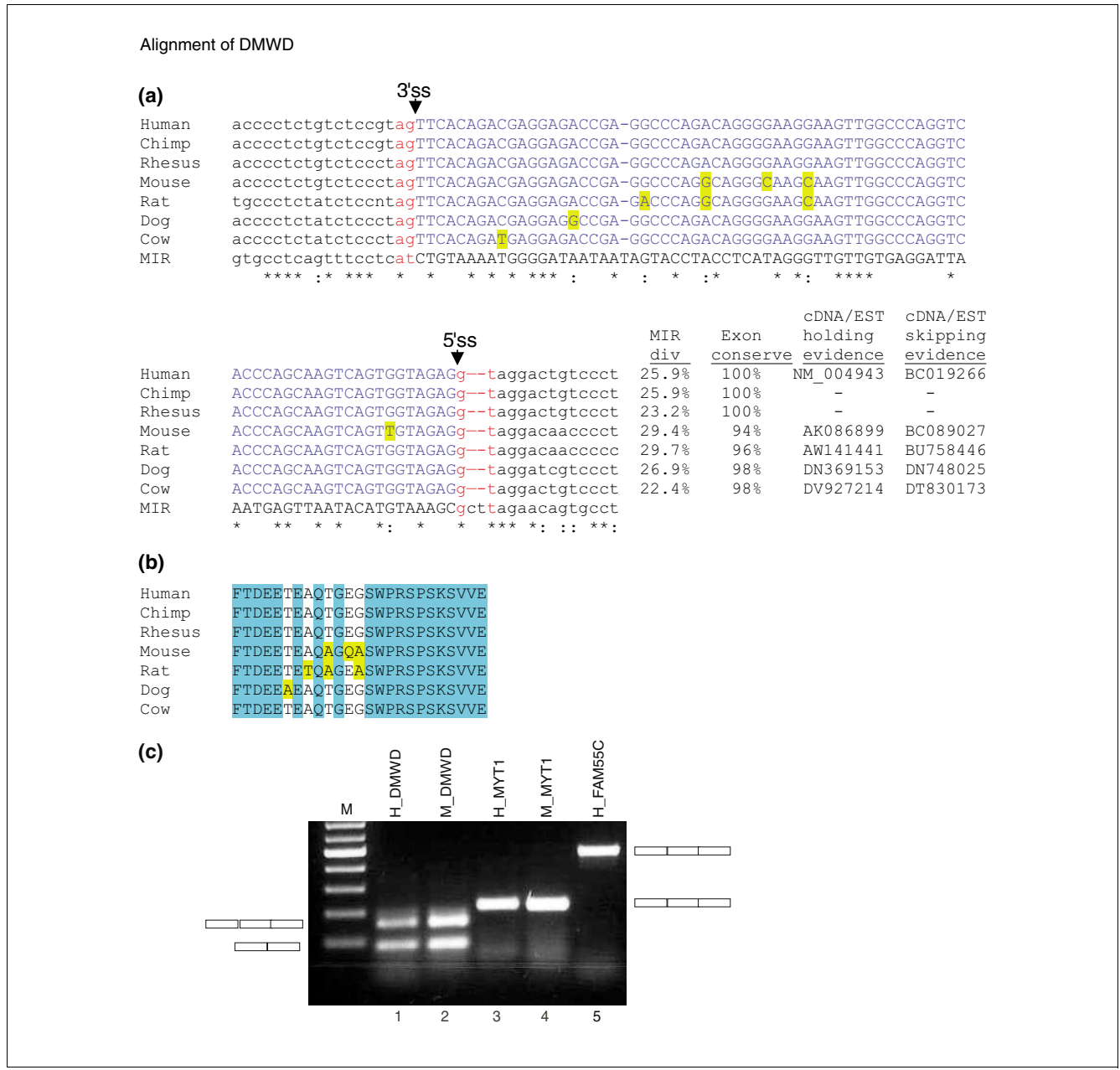

**Figure 2**

RT-PCR analysis of selected *Alu* and MIR exons. **(a)** Multiple alignment of mammalian interspersed repeat (MIR) exon in DMWD gene among mammals. Exon sequences are marked in blue, flanking intronic sequences are marked in black, and the canonical AG and GT dinucleotides at the 3'ss and 5'ss are marked in red. Nucleotide conservation is marked at the lower edge, with asterisks indicate full conservation and colons indicating partial conservation relative to the MIR consensus sequence (lower row). The divergence in percentage from the consensus MIR sequence is indicated under (MIR div); exon conservation in percentage compared with the human exon is indicated under (exon conserve); EST/cDNA accession confirming the exon insertion is indicated under (cDNA/EST holding evidence), and skipping is indicated under (cDNA/EST skipping evidence). Nonconserved nucleotides are marked in yellow. **(b)** This panel is similar to panel a, except that the conservation is shown for the protein coding sequence. **(c)** Total RNA was collected from SH-SY5Y human cell line and mouse brain tissue. Reverse transcription polymerase chain reaction (RT-PCR) analysis amplified the endogenous mRNA molecules using primers specific to the flanking exons. The PCR products were separated on an agarose gel, extracted and sequenced. A schema of the mRNA products is shown on the left and right. Columns 1 to 4 show the splicing pattern of orthologous human (H) and mouse (M) exons originating from the MIR element. Columns 1 and 2 show alternative splicing of an ortholog MIR element in both human and mouse, respectively (exon 4 in DMWD gene), and columns 3 and 4 show a constitutive pattern in both species (exon 5 in the MYT1L gene). Column 5 shows constitutive splicing of an *Alu* element in the human exon 3 of FAM55C gene. All PCR products were confirmed by sequencing. We cannot fully reject the option that an exon that is constitutively spliced under the above conditions is alternatively spliced in other cells or conditions. However, the constitutive selection is also supported by EST/cDNA coverage.

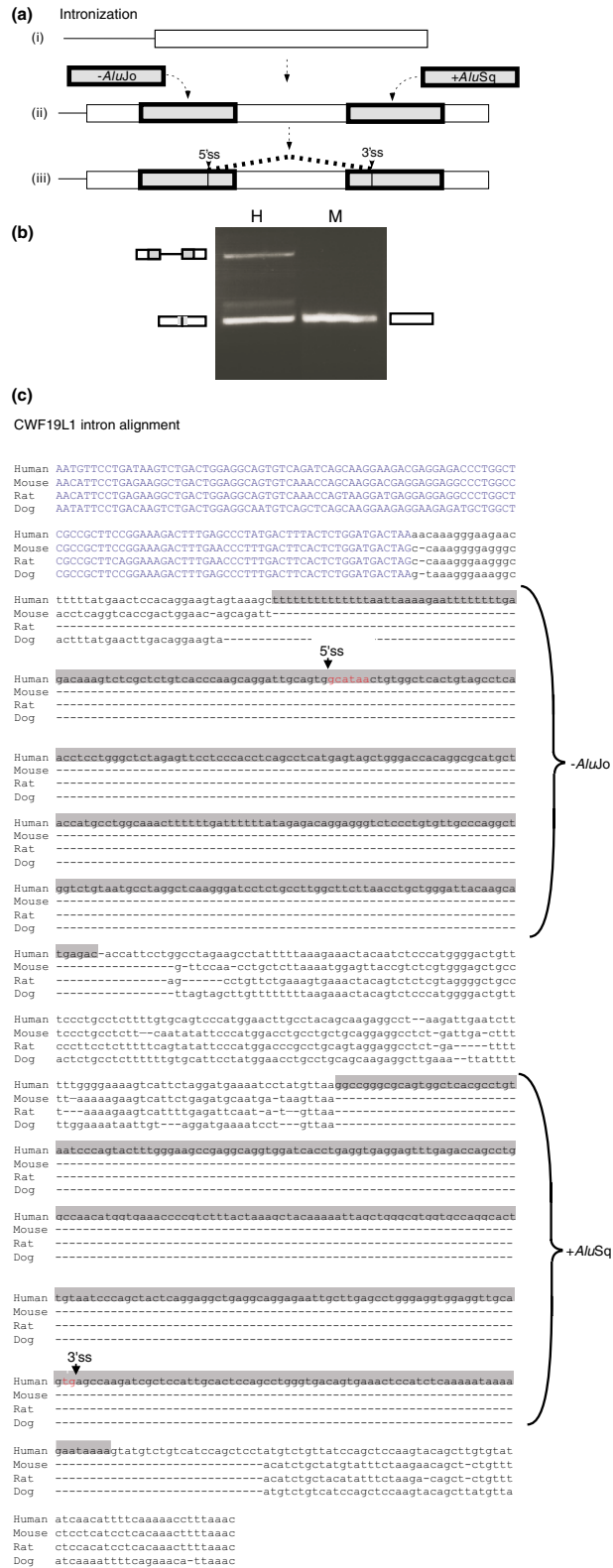

**Figure 3** *(see legend on next page)*

**Figure 3** *(see previous page)*

*Alu* insertions into an exon activate intronization in the CWF19L1 gene. **(a)** Intronization. (i) Illustration of the last exon of the CWF19L1 gene in mouse. (ii) During primate evolution, two *Alu* elements were inserted into the exon. (iii) Because of these insertions, an intronization process activates two splice sites within the exon, a 3' and a 5' splice site. The isoform in which the intron is spliced out is supported by 12 mRNA/expressed sequence tags (ESTs), and the isoform in which the intron is retained is supported by four mRNA/ESTs. **(b)** Testing the splicing pathway of this exon between human and mouse. Polymerase chain reaction (PCR) analysis on normal cDNAs from human kidney (marked H) and from mouse brain tissue (marked M). PCR products were amplified using species-specific primers, and splicing products were separated in 1.5% agarose gel and sequenced. **(c)** Alignment of the sequence of the last exon of the CWF19L1 gene among human, mouse, rat, and dog is shown. The two *Alu* elements are marked in gray. The selected 5'ss and 3'ss are marked.

respectively). The analysis of the number of TE insertions within the first or last exon in human and mouse was done on UCSC annotated genes, in which a consensus mRNA sequence exists. We searched for TE insertions within the first and last exon of 19,480 human and 16,776 mouse genes that are listed as known genes in the UCSC genome browser. In human annotated genes, the average length of the first and last exon is 464.6 base pairs) and 1,300 bp, respectively. In contrast, in mouse genes the first exon has an average length of 392.7 bp and the last exon an average length of 1,189 bp. Our analysis revealed that 3,686 TEs were inserted within the first and 10,541 TEs within the last exon of the human transcriptome. In the mouse transcriptome, 1,932 and 7,847 TEs were inserted into the first and last exons, respectively (Figure 1b). On average, the human transcriptome is significantly enriched with TEs: 3.5% and 7.6% of the first and last exons in human coding genes contain TE insertions, as compared with 0.4% and 1.7% of first and last exons in mouse coding genes that contain TE insertions (Mann-Whitney; first exon $P = 0$ and last exon $P = 0$). One-third of all TE insertions within the human first and last exons belong to *Alu* (35.3%), although *Alu* elements comprise only 27.9% of TEs within the human genome ($\chi^2$; $P < 10^{-9}$ [degrees of freedom = 1]). When normalizing for the differences in length of the first and last exons, there is no bias for TE insertion within either the first or the last exon of the gene.

### *Alu* element insertion generates new introns

We found four cases in which the insertion of the *Alu* element into the last exon of the gene was involved in the activation of an alternative intron (called intron retention) within the 3'-UTR of the gene (primate-specific intron gain events). Here, new splice sites were introduced within the last exon of the gene. These events occurred within the *SS18L1, PDZD7, C14orf111,* and *CWF19L1* genes (illustrated in Figure 1b [iv]).

In the *SS18L1* gene, in which the *Alu* was inserted in the sense orientation, three mutations within the *Alu* sequence activated a 5'ss, whereas the 3'ss and the polypyrimidine tract (PPT) was contributed from the conserved area of the exon. In the *CWF19L1* gene, the last exon is conserved within the mammalian class. Two *Alu*s were inserted into that exon, one in the sense orientation and the other in the antisense orientation, and the 5'ss and 3'ss were contributed by antisense *Alu*Jo and by the sense *Alu*Sx, respectively (shown in Figure 3a,c). Examination of the splicing pattern of this exon in human and mouse by RT-PCR revealed that the exon is con-stitutively spliced in mouse (Figure 3b, lane 3). However, in human, the same analysis on kidney normal tissue detected two RNA products: intron retention isoform (upper PCR products; Figure 3b, lanes 1 and 2) and spliced product using 3' and 5' spliced sites within the *Alu*s (Figure 3b, lane 1, lower RCR product). See Figure 3a for a graphical illustration of these splice sites and Figure 3c for their location along the exonic sequence. The spliced intron is flanked by a canonical 5'ss of the 'GC' type and a noncanonical 3'ss of 'tg' instead of 'ag' (see Figure 3c). The identity of these splice sites was confirmed by sequencing and was supported by 12 cDNA/EST as well, indicating that the same noncanonical splice site is used in all cases (for the list of these cDNA/ESTs, see Additional data file 8). We currently cannot explain how the splicing machinery selects a noncanonical splice site, although it was shown previously that a 'tg' spliced site can serve as a functional 3'ss [32,33]. Additionally, it may also be related to RNA editing, because of formation of dsRNA between the sense and antisense *Alu* (see, for example, the report by Lev-Maor and coworkers [16]). This hypothesis is supported by detection of potential deviation between the genomic sequence and some of the cDNA in the flanking exonic sequences. However, further analysis is needed to understand this phenomenon fully.

With regard to the last two genes exhibiting intronization, the *C14orf111* and *PDZD* genes, the last exon is not conserved within mammals. In the *C14orf111* gene the last exon comprises L1, three *Alu* elements, and an LTR insertion. The intron retention is spliced by a 3'ss and a 5'ss that are found within the *Alu* sequences (Genebank accession BC08600 and BX248271 confirm the splicing of the intron, and BX647810 confirm the unspliced intron). In the *PDZD* gene there were two *Alu* insertions. Both the 3'ss and the 5'ss are found within the *Alu* sequence (Genebank accession BC029054 confirm the splicing of the intron, and AK026862 confirm the unspliced intron). All of these cases are within the last exon of the gene, within the 3'-UTR. The intronizations generate an alternative intron, that is, both the *Alu* insertion and spliced forms are present in the mRNA.

### Short interspersed nuclear elements tend to exonize in the antisense orientation

Our dataset shows that *Alu* and MIR have a statistically significant bias toward exonization in their antisense orientation, relative to the direction of the mRNA in the human transcriptome. Additionally, B1, MIR, B2, and B4 are biased

**Table 3**

**Architecture of the newly recruited exons in the human genome**

| RE | Sense | Antisense | Whole | 5'ss | 3'ss |
|---|---|---|---|---|---|
| *Alu* | 139 (13%) | 921 (87%) | 701(66%) | 240 (23%) | 119 (11%) |
| MIR | 60 (33%) | 121 (67%) | 68 (38%) | 62 (34%) | 51 (28%) |
| L1 | 62 (28%) | 157 (72%) | 149 (68%) | 34 (16%) | 36 (16%) |
| L2 | 41 (40%) | 62 (60%) | 42 (41%) | 31 (30%) | 28 (29%) |
| CR1 | 4 (33%) | 8 (67%) | 6 (50%) | 3 (25%) | 3 (25%) |
| LTR | 68 (44%) | 87 (56%) | 103 (66%) | 19 (10%) | 33 (24%) |
| DNA | 47 (50%) | 46 (50%) | 46 (50%) | 22 (23%) | 25 (27%) |

The first column indicates the different transposed elements (TEs) that were examined. In columns 2 and 3, the numbers of exonizations in the sense and antisense orientations are shown. The percentages of the total number of exonizations are given in parenthesis. In columns 4, 5, and 6, the numbers of exons are given in which the TE contributes the whole exon, the 5', and the 3' part of an exon, respectively. In parentheses are given the percentage of the total number of exonizations. LTR, long terminal repeat; MIR, mammalian interspersed repeat; RE, retroelement.

**Table 4**

**Architecture of the newly recruited exons in the mouse genome**

| RE | Sense | Antisense | Whole | 5'ss | 3'ss |
|---|---|---|---|---|---|
| B1 | 34 (24%) | 108 (76%) | 58 (41%) | 55 (39%) | 29 (20%) |
| MIR | 5 (18%) | 23 (82%) | 12 (43%) | 8 (28.5%) | 8 (28.5%) |
| B2 | 23 (28%) | 60 (78%) | 35 (42%) | 31 (36%) | 19 (23%) |
| B4 | 20 (31%) | 45 (69%) | 26 (40%) | 17 (26%) | 23 (34%) |
| L1 | 47 (46%) | 56 (54%) | 77 (75%) | 15 (14.5%) | 12 (10.5%) |
| L2 | 1 (9%) | 10 (91%) | 3 (28%) | 4 (36%) | 4 (36%) |
| LTR | 35 (49%) | 37 (51%) | 48 (67%) | 10 (14%) | 14 (19%) |
| DNA | 6 (54%) | 5 (46%) | 4 (36%) | 4 (36%) | 3 (28%) |

The first column indicates the different transposed elements (TEs) that were examined. In columns 2 and 3, the numbers of exonizations in the sense and antisense orientations are shown. The percentages of the total number of exonizations are given in parenthesis. In columns 4, 5, and 6 are shown the numbers of exons are given in which the TE contributes the whole exon, the 5', or the 3' part of an exon, respectively. In parentheses, the percentages out of the total number of exonizations are given. LTR, long terminal repeat; MIR, mammalian interspersed repeat; RE, retroelement.

toward the antisense exonization in the mouse transcriptome (see Tables 3 and 4, columns 2 and 3). We correlate this phenomenon with the fact that, in most cases, SINE elements contain a polyA tail at the end of their sequence. In the antisense direction, this polyA becomes a polypyrimidine tract that facilitates exonization [4,5]. LINEs and DNA repeats in both human and mouse do not exhibit a preferential exonization orientation (the greater number of L1 exonizations in the antisense is caused by its biased insertion in the antisense direction within introns, and not because of a preferential exonization in the antisense orientation). LTRs exhibit a biased exonization in their sense orientation in both human and mouse (for χ² test *P* value, see Additional data file 3).

### *Alu*, L1, and long terminal repeat have the highest capability to contribute a whole exon

An exonization can occur if the TE contributes only a 5'ss or 3'ss to the exon or by using both intrinsic 5'ss and 3'ss within the TE (entire exon). We divided our TE exon dataset into three groups: those that contributed a whole exon and those that contributed only a 5'ss or only a 3'ss (Tables 3 and 4, columns 4 to 6, respectively). In 66% of exonized *Alu* and LTR and 68% of exonized L1 elements in the human transcriptome, the whole exon is contributed by the TE. In the mouse transcriptome, 75% of exonized L1 and 67% of exonized LTR are entire exons. In contrast, all other TE exonizations contribute a complete exon in approximately 40% of the cases, rates that are significantly lower than those for *Alu*, L1, and LTR (χ²; *P* < 10⁻³ [degrees of freedom = 6] for human and *P* = 0.05 [degrees of freedom = 5] for mouse). The reason for the high level of *Alu* exonization is the low number of mutations needed to activate potent splice sites [4,5], as well as the presence of enhancers and silencers that were previously reported to reside within the *Alu* consensus sequence [34]. This observation suggests that *Alu*, L1, and LTR TEs have greater potential to be recognized by the spliceosome machinery, and probably many copies of these TEs serve as 'pseudo-exons' (intronic *Alu* sequences containing putative 5'ss and polypyrimidine tract-3'ss that are one mutation away from exonization) within introns of protein coding genes [4,5].

**Table 5**

**The effects of exonization on human protein coding regions**

| RE | A Alt. CDS | B CDS | C UTR | D Stop | E Frameshift | F Functional |
|---|---|---|---|---|---|---|
| *Alu* | 15 (1.5%) | 650 (61.5%) | 396 (37%) | 399 (61%) | 158 (24%) | 93 (14%) |
| MIR | 3 (2%) | 100 (55%) | 78 (43%) | 84 (84%) | 9 (9%) | 7 (7%) |
| L1 | 4 (2%) | 144 (66%) | 71 (32%) | 107 (74%) | 21 (15%) | 16 (11%) |
| L2 | 2 (2%) | 66 (64%) | 35 (34%) | 50 (76%) | 8 (12%) | 8 (12%) |
| LTR | 9 (6%) | 76 (49%) | 70 (45%) | 46 (60%) | 14 (19%) | 16 (21%) |
| DNA | 0 (0%) | 77 (81%) | 18 (19%) | 53 (69%) | 20 (26%) | 4 (5%) |

The first column shows the different examined transposed elements (TEs). Columns 2, 3, and 4 show the positions of TE exonization within the mRNA: creating an alternative coding sequence (CDS) start (Alt. CDS), exonization within the CDS, or exonization within the untranslated region (UTR). The relative percentages are given in parentheses. Columns 5, 6, and 7 show the effect of exonization within the CDS: exonizations that contain an in-frame stop codon (within the exon); exonizations that create a frameshift in the CDS but do not contain an in-frame stop codon; and functional exonizations (exons that do not possess an in-frame stop codon and do not cause frameshifts). The relative percentages within CDS are indicated in parentheses. The total number of TEs (100%) is found at the foot of the second column of Table 1. LTR, long terminal repeat; MIR, mammalian interspersed repeat; RE, retroelement.

## Do transposed element exonizations enter with the same probability in all parts of the mRNA?

A new exon resulting from TE exonization can reside either within the CDS or the UTR. When inserted within the UTR, the exon can introduce an alternative start-of-coding sequence or it can enlarge the UTR. The different number of exonizations within the mRNA for different TEs is summarized in Tables 5 and 6 (columns 2 to 4) for human and mouse data, respectively. More than 32% of all exonized TEs in both human and mouse are inserted within the UTR regions. To check whether exonization has a bias toward insertion in the UTR or the CDS, we estimated the fraction of the UTR and CDS within human and mouse genes, based on the annotations of 19,480 human and 16,776 mouse genes, respectively (see Materials and methods, below). In human, the average gene length is 59,186 nucleotides, in which 79% and 21% are CDS and UTR sequences. In mouse, the average gene length is 49,101, in which 73% and 27% are CDS and UTR sequences. Our results revealed a statistically significant bias for exonization of new TE exons in the UTR, as compared with CDS regions, for *Alu*, MIR, L1, and L2 in human and for B1, MIR, B2, B4, L1, and L2 in mouse (for $\chi^2$ test *P* values, see Additional data file 4). This UTR bias is probably related to selection against exonization in the CDS.

## How many transposed element exons potentially contribute to proteome diversity?

The majority of exonizations in our dataset inserted an in-frame stop codon within the CDS (in 61% to 84% of the cases); in 9% to 24% of the cases they caused a frame shift in the reading frame. Therefore, between 81% and 93% of the exons were potentially harmful because they produced a truncated protein. Only a small fraction of between 7% and 19% did not possess an in-frame stop codon and did not generate a frame

**Table 6**

**The effects of exonization on mouse protein coding regions**

| RE | Alt. CDS | CDS | UTR | Stop | Frameshift | Functional |
|---|---|---|---|---|---|---|
| B1 | 4 (3%) | 87 (65%) | 43 (32%) | 67 (77%) | 13 (15%) | 7 (8%) |
| MIR | 3 (11%) | 9 (33%) | 15 (56%) | 5 (55%) | 0 (0%) | 4 (45%) |
| B2 | 4 (5%) | 37 (46%) | 40 (49%) | 29 (78%) | 3 (8%) | 5 (14%) |
| B4 | 3 (5%) | 39 (63%) | 20 (32%) | 30 (77%) | 5 (13%) | 4 (10%) |
| L1 | 3 (3%) | 54 (53%) | 45 (44%) | 38 (70%) | 6 (11%) | 10 (19%) |
| L2 | 0 (0%) | 3 (33%) | 6 (66%) | 1 (33%) | 1 (33%) | 1 (33%) |
| LTR | 3 (4%) | 38 (53%) | 31 (43%) | 29 (76%) | 6 (16%) | 3 (8%) |
| DNA | 1 (1%) | 4 (36%) | 6 (67%) | 3 (75%) | 0 (0%) | 1 (25%) |

The first column shows the different examined transposed elements (TEs). Columns 2, 3, and 4 show the positions of TE exonization within the mRNA: creating an alternative coding sequence (CDS) start (Alt. CDS), exonization within the CDS, or exonization within the untranslated region (UTR). In parentheses, the relative percentages are given. Columns 5, 6, and 7 show the effect of exonization within the CDS: exonizations that contain an in-frame stop codon (within the exon); exonizations that create a frameshift in the CDS but do not contain an in-frame stop codon; and functional exonizations (exons that do not possess an in-frame stop codon and do not cause frameshifts). The relative percentages within CDS are indicated in parentheses. The total number of TEs (100%) is found at the foot of the second column of Table 2. LTR, long terminal repeat; MIR, mammalian interspersed repeat; RE, retroelement.

shift, and thus potentially contributed a new function to an existing protein (Tables 5 and 6, columns 5 to 7). When these exons were searched against PROSITE [35,36], 54 out of the 93 *Alu* exons, three out of seven MIR exons, six out of 16 L1 exons, five out of 8 L2 exons, and none of 17 LTR and DNA exons were found to add a new protein domain (Additional data file 9). Overall, 68 exons out of 141 (48%) exhibited a hit against a domain in PROSITE [36], reducing the number of domain-contributing TE exonizations to 4.3% (in mouse, only one hit against a domain in PROSITE was found). Thus, our results show that a small fraction of relatively young exonized TEs has the potential to contribute to protein functionality. However, we cannot rule out the possibility that the TE exons that do not add a new protein domain also contribute to proteome complexity by inserting into an existing protein domain. Such is the case for exon 8, which is an *Alu* exonization within ADAR2; the *Alu* exonization that was inserted into the deaminase domain creates a twofold difference in this gene's specific editing activity [12].

### Do new exons resulting from transposed element exonizations differ in their characteristics from conserved alternatively spliced cassette exons?

We next examined the characteristics of these new exons resulting from TE exonization. Conserved alternatively spliced exons are under selective forces different from those in constitutively spliced exons [11,37]. These exons contain weaker 5'ss (ΔG), are shorter than constitutively spliced exons, and have a high inclusion level with respect to the new *Alu* exons [3]. Therefore, we examined these characteristics among the different exons that originated from TEs and compared the findings with those for 596 and 44,732 alternatively skipped exons and constitutive exons conserved between human and mouse, respectively.

The TE exons have a low inclusion level, with an average of 19.17 ± 26.2% in human and 26.51 ± 31.8% in mouse, the inclusion level of human TEs being significantly lower (Mann-Whitney; $P < 10^{-6}$; Figure 4c). Both values are significantly lower than the 64.39 ± 31.1% of conserved alternatively spliced exons (Mann-Whitney; $P = 0$ and $P < 10^{-66}$, respectively). The TE exons are, on average, 143.4 ± 118.4 bp long in human and 133.6 ± 75.1 bp in mouse (Figure 4d). They are therefore significantly longer than conserved alternative exons, in which the average length is 97.7 ± 56.7 nucleotides (Mann-Whitney; $P < 10^{-38}$ and $P = 0$, respectively), and similar to conserved constitutive exons in which the average is 132.4 ± 49.9 nucleotides (Mann-Whitney; $P = 0.7$ and $P = 0.8$, respectively). In addition, the TE exons have a very weak 5'ss, relative to alternatively spliced exons in which the average U1/5'ss strength (ΔG) is -4.87 ± 2.26 kcal/mol for human exons and -4.88 ± 2.26 kcal/mol for mouse (Figure 4a). This in turn is significantly weaker than the conserved alternative exons, whose ΔG is -5.62 ± 1.9 kcal/mol (Mann-Whitney; $P < 10^{-9}$ and $P < 10^{-6}$, respectively). Conserved alternatively spliced exons have already been shown to have a significantly weaker U1/5'ss strength than constitutively spliced exons [11]. In humans, the TE exons that originated from *Alu* have the lowest inclusion level, the weakest 5'ss and the shortest exons, as compared with all other exonized TEs. In addition, exonized *Alu*s have low divergence from the consensus sequence, meaning that not many mutations are needed for their exonization. In contrast, the MIR exons have the strongest 5'ss in both human and mouse, the highest inclusion level, and these exons are also the most diverged exons among SINEs (with respect to the consensus sequence) in both human and mouse (Figure 4b). The high inclusion level could be explained by the fact that the MIR element has one major 5'ss (Figure 5c), that contains an almost canonical 5'ss sequence (CTA/gtaagt). This is also consistent with the finding that the MIR exons contribute the highest level of constitutive exonization in both human and mouse (Figure 1a [v]) and have a relatively high inclusion level (Figure 4c).

## Discussion
### The majority of transposed elements have a biased insertion in introns
Our analysis revealed that the majority of TEs reside in intronic sequences, although introns comprise only 24% of the human genome [1]. SINEs tend to be localized in GC-rich regions, which reflect gene-rich areas [1]. The insertions into introns presumably reflect selection against insertion of TEs into protein coding exons and preferable retroposition into transcribed regions. L1 is the only active autonomous non-LTR retrotransposon in the human and mouse genome. In contrast to SINE elements, L1 tends to reside within AT-rich areas in both human and mouse [1,2]. It is assumed that L1 elements, as well as LTRs, with their larger size in open reading frames and the presence of polyadenylation signals, may be more deleterious within genes than smaller SINE elements [38]. This hypothesis is supported by the bias toward insertion in the antisense orientation of both L1 and LTR within the human and mouse introns. Furthermore, L1 insertions within human introns were shown to reduce gene expression [39-41], and inactive LINE-1 can slow down transcription. In agreement with this hypothesis, L1 in both human and mouse has the lowest presence within introns. The older, nonactive LINEs, L2, and CR1 (L3), and DNA repeats have the same percentage of presence within introns as the shorter SINE elements, probably because of their inactive state, which is tolerated by gene-rich areas.

### Do transposed element exons reflect splicing errors?
The large fraction of exons within our dataset for which there is only one piece of supporting EST/cDNA evidence raises an intriguing question as to whether these exons reflect temporary splicing mistakes. Probably, some of these exons are indeed splicing errors, reflecting the low fidelity of the spliceosome machinery [42,43]. Another explanation is that the new exon added to a gene will first have a very weak splice signal, and therefore it will be included in only a small fraction of mRNA (approximately 10%). As a consequence, this addition is free to evolve; with time, positive selection can strengthen its splice sites, resulting in an increase in its fraction within the transcriptome [3,14,44]. Our findings are in agreement with this hypothesis. The *Alu* elements with the lowest inclusion level are the most recent insertions within the human genome, and MIR elements with the highest inclusion level are much older. These 'newcomers', the low inclusion exons, are the sites for future possible exaptation and fixation within the human transcriptome. Presumably, these exons are subjected to selective pressures that determine their future evolution in terms of improved recognition by the splicing machinery, as well as protein fixation, or alternatively complete loss [10]. Long evolutionary periods are needed for successful exaptation events, and these new exons are the potential raw materials for future evolution in human and mouse [3,6,7,44].

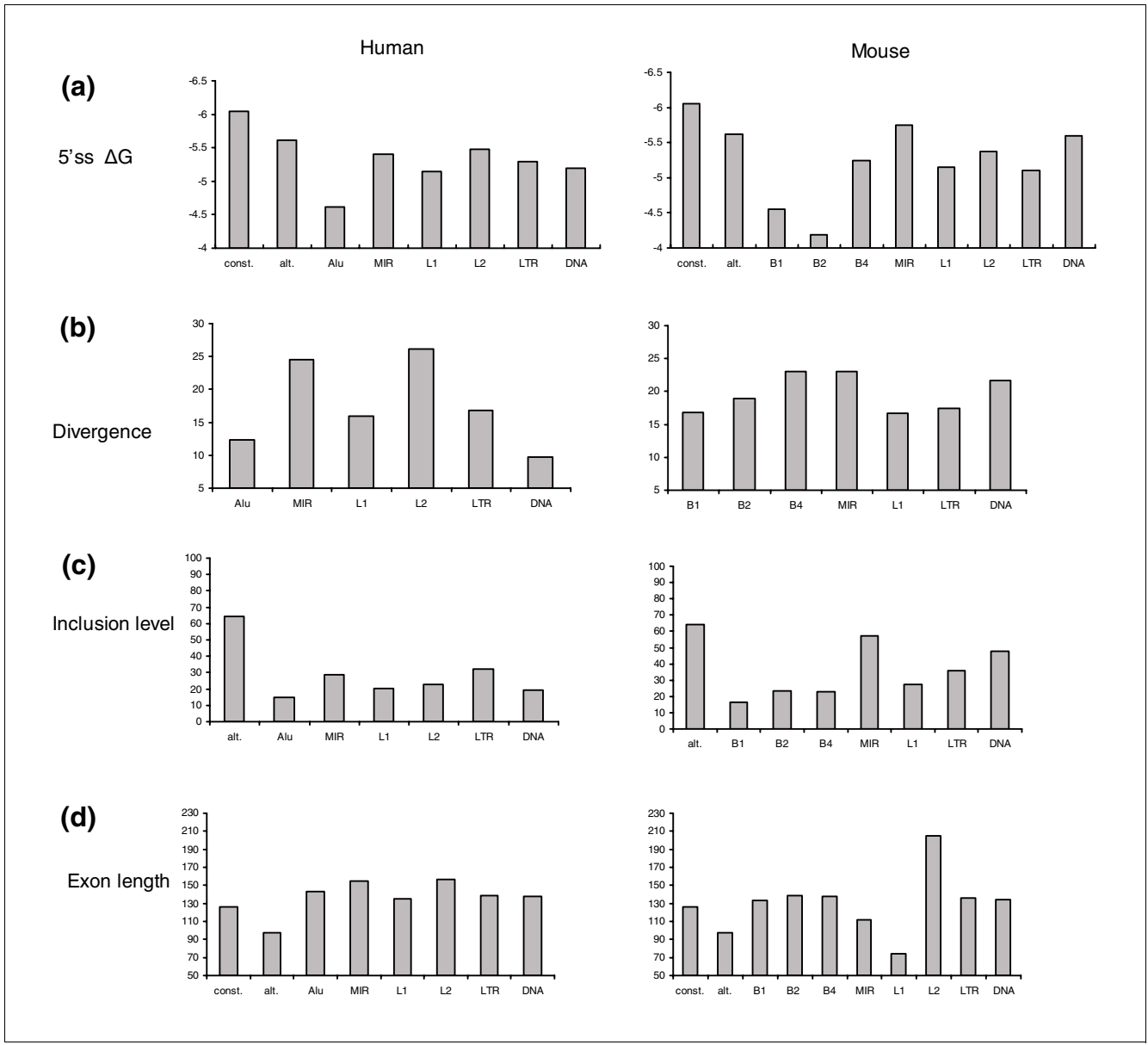

**Figure 4**
Characteristics of the exonized TEs. **(a)** The free energy resulting from the base pairing of the 5'ss with U1 snRNA (ΔG) for the indicated exonized transposed elements (TEs). 'const.' and 'alt.' indicate conserved human and mouse constitutively spliced and alternatively spliced exons, respectively. **(b)** The divergence in percentage from the consensus sequence. **(c)** The average inclusion level. **(d)** The average exon length.

## Higher transposed element exonization levels in human due to *Alu* insertions
Most of the new exons are exonized from intronic sequences. In rodents, these new alternatively spliced exons originated from unique intronic sequences [45]. However, in the human genome most of those exons came from highly repeated sequences (in which 40% are *Alu* exons) [44]. An explanation for this discrepancy might be that the main source for these differences is the higher level of *Alu* elements within human introns, along with its extraordinary high level of exonization.

Why is it that *Alu* elements are so able to exonize? Both elements, *Alu* in human and B1 in mouse, have the same ancestral origin, namely the 7SL RNA, but they still differ tremendously in their ability to exonize. This could be explained by the *Alu* element being a dimer and B1 a monomer. Therefore, looking at the *Alu* as a double B1, we would expect twice as many exonizations of *Alus* as compared with B1 elements. However, the exonization level of the *Alu* element is almost three times higher than that of B1 (Fisher's exact test [$\chi^2$] between a double amount of exonization of B1 and the *Alu* actual exonization level gives a statistically significant difference; $P < 10^{-11}$). Thus, it seems that the dimeric structure of *Alu* has a synergetic effect on alternative exonization. *Alu* comprises two arms, with the right arm having an addition of 31 nucleotides with respect to the left arm. B1 is homologous to the left arm of the *Alu*. The majority of *Alu* exonizations occur within the right arm; 959 of the exons are found within the right arm, and only 24 are found within the left arm. Both B1 and the left arm of the *Alu* have the same highly selected 5'ss and 3'ss (Figure 5a,b). However, the right arm of the *Alu* contains more potential splice sites (four 5'ss and three 3'ss). This indicates that the *Alu* dimeric structure, along with its unique left arm sequence that does not exist in B1, are the main contributors to *Alu*'s extraordinary ability to exonize.

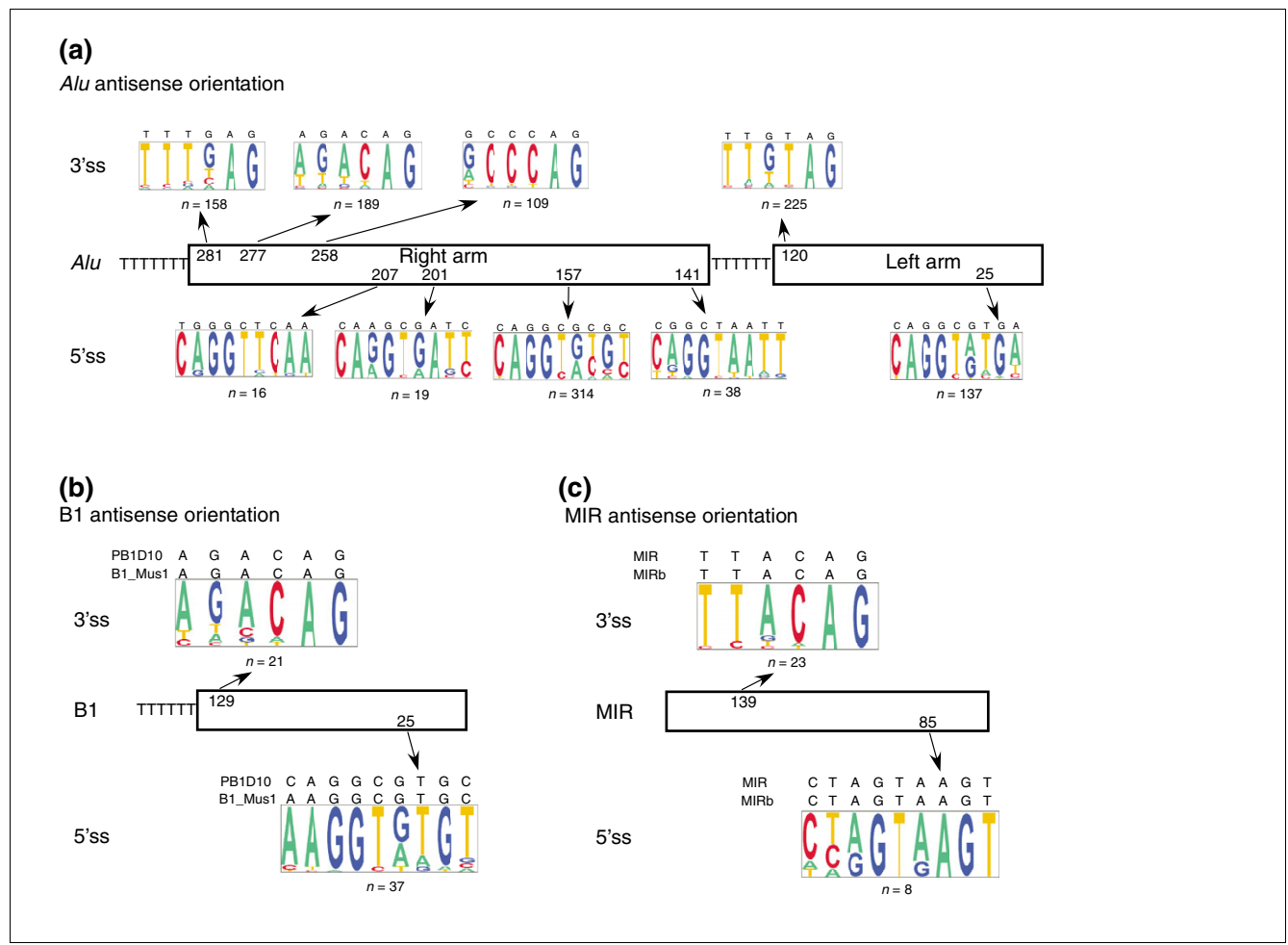

**Figure 5**
5' and 3' splice site selection of *Alu*, B1, and MIR. Both *Alu* and B1 originated from 7SL RNA. *Alu* has a dimeric form with a very similar left and right arm, whereas B1 has a monomeric form similar to the left arm of the *Alu* element. **(a)** The most selected 5'ss and 3'ss within the exonized *Alu* element in the antisense orientation. The right and left arm are shown by boxes. The numbers within the boxes indicate positions (according to the *Alu* consensus sequence) in which the prevalent 3'ss and 5'ss are selected, above and below the boxes, respectively. Pictogram depiction of each splice site is shown, the consensus sequence of *Alu* is marked above the pictogram, and the number of times that the site was selected is shown below it. **(b and c)** Similar presentation as in panel a for the prevalent 5'ss and 3'ss selected in B1 and mammalian interspersed repeat (MIR) elements, respectively.

## Potential splice sites contributed by transposed elements

All TEs contribute many potential splice sites both in their sense orientation and in their antisense orientation (for numbers, see Additional data file 11). All SINEs have substantially more splice sites in their antisense orientation, accounting for their preferential exonization in the antisense orientation. In addition, all SINEs have a polyA tail at the end of their sequences. This polyA tail becomes a polyT in the antisense orientation. This sequence can later serve as a polypyrimidine tract and facilitate exonization, as was previously shown for *Alu* elements' 3'ss selection [4]. The 'AG' downstream of the polyT is also the most selected 3'ss in all SINEs (as can be seen for B1, *Alu*, and MIR in Figure 5).

## *Alu* insertion gave rise to alternatively spliced human introns

The mechanism by which an insertion of a SINE element gave rise to a new intron in the coding region of the catalase A gene of rice was previously reported [21]. Here we show that this mechanism is not limited to plants and is present in the human genome also. Thus, the mechanism by which new introns are generated by a SINE insertion also exists in the human genome, leading to the conclusion that this intron gain mechanism presumably is widespread outside plant genomes as well. The new intron is within the coding sequence, but all four new introns reported here are within the 3'-UTR. A noncanonical 3'ss is used by the intron created within the CWF19L1; this intron is spliced at a 'tg' signal, although there are two flanking 'ag' signals that could be used as a canonical 3'ss. The use of a noncanonical 3'ss was previously reported [32]. We can also explain the use of the noncanonical 3'ss by the possible secondary structure that is created as a result of the presence of two *Alus* in close proximity and in opposite orientation, creating a presumably dsRNA form, which may disrupt the correct splicing of this intron. Evidence for the secondary structure is given by various A-I editing sites that can be seen within these *Alus*. This structure of two *Alus* in reverse orientation was shown to be the target of editing [16]. Although such intronizations were not found in mouse, we expect that in the metazoan such events are more abundant within the UTRs than in the CDS, because such events create a substantial rearrangement that have a deleterious effect on the protein coding sequence. Previous reports failed to detect intron gain within the mammalian class [23,24], because these reports did not analyze alternatively spliced introns and nonconserved regions within human genes.

## Transposed element exonization is a source of newly constitutively spliced exons

Our findings show that TE exonization mostly introduces a low inclusion exon-skipping event (also see [3,44]). Experimental analysis of selected exons has shown that TE exonization can introduce a constitutively spliced exon within the tested tissue and under the experimental conditions of the RT-PCR experiments. This event can be the result of two possible scenarios: a newly constitutive exon either emerged in one or in two steps, in which an alternative selection precedes a constitutive one. The existence of many diseases created by the introduction of a *de novo Alu* constitutive exonization, along with a small number of mutations that were shown to create a constitutively spliced *Alu* [11], supports the former scenario. The significantly higher percentage of constitutively spliced MIR (which is a more ancient TE within the human genome) with respect to *Alu* ($\chi^2$; $P < 0.003$ [degrees of freedom = 1]) supports the latter option. Probably, these two scenarios reside side by side.

Our findings indicate that most of the exonizations generate alternatively spliced exons. However, a small portion of the exonizations generates constitutively spliced exons. The question that arises is why new exons that are constitutively spliced are not selected against? This is because they are mostly found within the UTRs (71% versus 37% of TE exonizations that are alternatively spliced), and when the exonizations occur within the CDS the exon lengths are mostly 'divisible-by-3', and therefore they do not interrupt the protein coding frame. In all other cases in which the exonizations were constitutive and not symmetric, they were introduced within hypothetical proteins. In these cases, these are either not *bona fide* protein coding genes, or they are fast evolving genes not bound to high selective pressure.

## *Alu* and L1 act as pseudo-exons

We found that approximately 65% to 68% of the exonizations originating from *Alu*, L1, and LTR are entirely contributed by the *Alu*, L1, and LTR sequences, respectively, as compared with approximately 40% contribution from all other TEs in both human and mouse. These findings suggest that *Alu*, L1, and LTR have the greatest potential for being recognized as an exon without assistance from additional signals found randomly within adjacent intronic sequences. This could explain the reported interference with the selection of abnormal splice sites caused by *Alu* and L1 insertion within introns (reviewed by [46]). The deleterious effect may be due to an insertion of a competitive target for the spliceosome machinery. We found about 750,000 intronic Alu elements, suggesting that insertion into introns of *Alus* is generally tolerated. However, several diseases caused by *de novo Alu* insertion within introns have been reported [46]. This implies that retropositions into functionally important sequences within introns, or *Alu* insertions that effect essential splicing activities, are selected against. Also, intronic *Alus* with a putative polypyrimidine tract and a downstream potential 5'ss may function as pseudo-exons that are not spliced, but compete for the binding of splicing factors. Such competition may alter mRNA splicing of the flanking exons, such as shifting the splicing of the surrounding exons from constitutive to alternative [46] (Lev-Maor G and coworkers, unpublished data); 7,810 intronic pseudo-*Alu* exons were identified previously [5].

## Transposed element insertions in the first and last exons

The UTR contains motifs that can regulate many aspects of mRNA function, such as nuclear export, cytoplasmatic localization, translational efficiency, and stability [47]. Moreover, alternative UTRs were shown to determine tissue-specific functions [48]. Our findings show that, on average, 3.5% and 7.6% of the first and last exons in the human genome contain TEs, as compared with only 0.4% and 1.7% of first and last exons in the mouse genome. These findings suggest a much greater effect of TE insertions into the UTRs of the human transcriptome as compared with the mouse transcriptome. However, we found no biased insertion of TEs in the first or last exon.

## How many transposed elements contribute to proteome diversity?

Our findings suggest that the potential contribution of exons originated from TEs to proteome complexity is very low. Only 4.3% of the exonized TEs potentially contributed a new functionality to the proteome, which is consistent with previous reports [3,7]. Gotea and Makalovski [7] argue that a long evolutionary period is needed for a successful exaptation event, and that it is unlikely that young TEs will contribute a new function to a protein. Our results are in agreement with this observation, indicating that although many insertion events happened, the fraction of potential functional TE exons is low (about 4%). However, the fraction of functional exonized TEs may be even greater than that, because some of the exonized TEs may affect other regulatory pathways, such as regulation of mRNA level by activation of nonsense-mediated decay (NMD), especially if the exon is in the 3' half of the gene [49-51]. Other studies suggest that not all of the isoforms that contain premature termination codon (PTC) are subject to degradation by NMD [52], and NMD might only reduce the amount of the PTC containing isoform [53]. Therefore, insertion of PTC-containing exons could potentially be a source of novel sequences that advance regulatory networks.

## Do transposed element exonizations lead to tissue or cancer-specific isoforms?

We found 74 (< 8%) exons in the human and 18 (< 4%) exons in the mouse genome that have a potential tissue-specific association. In addition, six genes were found in which the *Alu* exonization was potentially specific to cancer. Two of them, YY1AP and KUB3, were reported to be involved in tumorigenesis [54-56].

## Conclusion

Our results demonstrate the importance of TEs in shaping both the human and the mouse transcriptomes in many different ways. However, the effect of TEs on the human transcriptome is several times greater than the effect on the mouse transcriptome, mostly because of the contribution of the primate-specific *Alu* elements.

## Materials and methods

### Dataset of transposed element exons in human and mouse genomes

The human NCBI 35 (hg17; May 2004) and the mouse NCBI33m (mm6; March 2005) assembly were downloaded, along with their annotations, from the UCSC genome browser database [28]. Coordinates of the EST and cDNA mapping were obtained from chrN_intronEST and chrN_mrna tables, respectively. TE mapping data were obtained from chrN_rmsk tables. A TE was considered to be intragenic if there was no overlap with ESTs or cDNA alignments; it was considered intronic if it was found within an alignment of an EST or cDNA in their intronic region. Finally, a TE was considered exonic if it was found within an exonic part of the EST or cDNA, if it possessed canonical splice sites, and if it was not the first or last exon of the EST/cDNA.

The insertions of TEs within EST/cDNA alignments were separated into two parts: those that entered within protein coding genes relative to the list included in the knownGene table in the UCSC genome browser [28] (based on proteins from SWISS-PROT, TrEMBL, TrEMBL-NEW, and their corresponding mRNA from GenBank), and other insertions within cDNA/ESTs alignments that were not mapped to the known genes list, and therefore were considered to be non-protein-coding genes. Non-protein-coding genes were defined as genomic regions covered by at least two correctly spliced cDNA/ESTs (flanked by canonical splice sites) containing at least three exons that did not overlap any annotated gene based on UCSC known genes lists, versions hg17 and mm6 for human and mouse, respectively. Unspliced genes were not included in our analysis; we only considered genes with at least two introns. Internal UTR exons were considered to be internal based on the annotations of knownGenes in UCSC and the fact that they were internal in the cDNA/EST. The TE position within the gene (UTR or CDS) and the exon phase were calculated based on the knownGenes table annotations of the gene start and end positions, as well as CDS start and end positions.

Splice site analysis was the same as that report by Sorek and coworkers [5]. For TE analysis, we used RepeatMasker [57] and Repbase annotations [58].

### Analysis of retroelement insertions within the first and last exons and assessment of untranslated region fraction in known genes

The tables knownGenes and kgXref were used to assess the relative lengths of the UTR and the CDS within 16,776 known mouse genes and 19,480 known human genes, as well as to find the first and last exons and to check for TE content. We have found that the UTRs comprise 21% of human genes and 27% of mouse genes, respectively, and the CDSs comprise 79% of human genes and 73% of mouse genes, respectively. These numbers served as the null hypothesis for comparison with the fraction of exonizations within the UTR and the CDS of all TEs. A goodness-of-fit $\chi^2$ test was used to asses the exonization fraction in every section of the gene.

## Statistical analysis

For the comparative analysis of exonization level, we used a contingency table $\chi^2$ test. When the contingency table was a 2 × 2 table, the Fisher's exact test was used. To assess the tendency of exonization within the UTR, we used the goodness-of-fit $\chi^2$ test. The null hypothesis was the fraction of the UTR and CDS within the known gene list of human and mouse (the calculation of this fraction is explained above).

## Calculation of exonization level and inclusion level

The definition of 'level of exonization' (LE) is the percentage of TEs that exonized ($N_E$) within the number of TE within introns ($N_I$):

$$LE = \frac{N_E}{N_I} \times 100$$

The definition of 'inclusion level' (IL) is the number of transcripts (cDNA/EST) that contain the exon (Nc) divided by the sum of the transcripts that include the exon (Nc) and the number of transcripts in which the exon is skipped (Ns).

$$IL = \frac{Nc}{Nc + Ns}$$

## Tissue classification of expressed sequence tags/cDNA and statistical analysis of tissue-specific and cancer-specific exons

The tissue source and the cancer/normal source of each EST were extracted from UniGene [59] or from GenBank annotations. We used Bayesian statistics as proposed in [30,31]. The criteria for high confidence of tissue specificity were TS > 50, rTS > 0.9, and rTS~> 0.9 (for details see [30,31]). A necessary condition for tissue specificity was at least three EST observations of the mRNA containing the exon in tissue T. For cancer specificity, an LOD score was calculated; only LOD scores above 2 (equivalent to $P < 0.01$) were considered to indicate cancer specificity.

## RT-PCR analysis of *Alu* and MIR exonization

All RNA was extracted from both the SH-SY5Y human cell line and mouse brain tissue. RT-PCR was performed using species-specific primers. Splicing products were separated on 1.5% agarose gel and confirmed by sequencing.

## RT-PCR analysis of *Alu* intronization in CWF19L1 gene

The spliced cDNA products derived from commercial cDNAs (BioChain) were detected by PCR, using endogenous forward and reverse primers: forward (human), GAGGTCCT-GGCCAGTGAAGCCA; reverse (human), GTACTTGGAGCT-GGATAACAG; forward (mouse) GAGGTCCTGGCCAGCGAAGCTA; and reverse (mouse) GTTCTTAGAAATACATAGCAG. Amplification was performed for 30 cycles, consisting of 1 min at 94°C, 45 s at 55°C, and 1.5 min at 72°C. The products were resolved on 1.5% agarose gel and confirmed by sequencing.

## Analysis of potential splice sites

The consensus sequences of the most abundant TE family from each TE class were analyzed for the existence of splice sites in both the sense and antisense orientations. The 5'ss was searched as 'gtnngn' or 'gcangn' (because these were shown to be active splice sites [5]). The 3'ss was searched as having at least 6 base pairs within its polypyrimidine tract [60] and a nearby (a distance of 3 base pairs at most) 'ag'.

## Additional data files

The following additional data are available with the online version of this paper. Additional data file 1 is a table describing the number of exonizations based on the existence of two ESTs/cDNA. Additional data file 2 is a table of all $\chi^2$ test $P$ values of all TE exonization levels. Additional data file 3 is a table of statistical $\chi^2$ test $P$ values for the preference of all TE exonizations in the sense/antisense orientation. Additional data file 4 is a table of statistical $\chi^2$ test $P$ values for the preference of all TE exonizations in the UTR. Additional data file 5 is an example of *Alu* exonization within a non-protein-coding gene. Additional data file 6 shows alignment of mouse L1 (Lx8) and human L1 (L1MC4). Additional data file 7 shows the populations of different families of L1 within human and mouse. Additional data file 8 is a table of cDNA and EST accessions, confirming the noncanonical 3' splice site of the alternative intron within CWF19L1 gene. Additional data file 9 is a table showing the domains contributed by TE exons. Additional data file 10 is a table of tissue and cancer-specific TEs. Additional data file 11 is a table of the potential splice sites of all TEs. Additional data file 12 shows RT-PCR of ACAD9 *Alu* exonization in different human cell lines. Additional data file 13 contains the coordinates of all tissue-specific and cancer specific exons in both human and mouse.

## Acknowledgements

We thank Hadas Keren, Schraga Schwartz, Eddo Kim, and Yair Bar-Haim for critical reading of the manuscript. This work was supported by the Cooperation Program in Cancer Research of the Deutsches Krebsforschungszentrum (DKFZ) and Israel's Ministry of Science and Technology (MOST), by a grant from the Israel Science Foundation (1449/04 and 40/05), MOP Germany-Israel, GIF, ICA through the Ber-Lehmsdorf Memorial Fund, and DIP. NS is funded in part by EURASNET. BM was supported by the grant Ca119 of the German-Israeli Cooperation in Cancer Research (DKFZ/MOST).